# Graphene with the secondary amine-terminated zigzag edge as a line electron emitter

Weiliang Wang and Zhibing Li[*]

State Key Laboratory of Optoelectronic Materials and Technologies

School of Physics and Engineering, Sun Yat-sen University, 510275 Guangzhou, P.R. China

ABSTRACT: An extraordinary low vacuum barrier height of 2.30 eV has been found on the zigzag-edge of graphene terminated with the secondary amine via the *ab initio* calculation. This edge structure has a flat band of edge states attached to the gamma point where the transversal kinetic energy is vanishing. We show that the field electron emission is dominated by the flat band. The edge states pin the Fermi level to a constant, leading to an extremely narrow emission energy width. The graphene with such edge is a promising line field electron emitter that can produce highly coherent emission current.

## I. INTRODUCTION

The graphene is a novel field electron emitter of which the emission in the edge direction can be uniform in large scale. Several groups have demonstrated that graphene electron emission assisted with moderate electric fields can be observable.[1-9] Edge properties are crucial for cold field electron emission (CFE). The characteristic feature directly relevant to CFE is the state-depending effective vacuum barrier height (EVBH) $W_p = W_0 - E_s + E_p$ with $W_0$ the local work function on the edge, $E_s$ the state energy relative to the Fermi energy, and $E_p$ the kinetic energy of motion parallel to the edge. There have been many studies on the peculiar edge properties of graphene.[10-15] The zigzag edge (Z-edge) being one of the most studied edges, can sustain edge states and resonances that are not present in other edges. It had been found that the Z-edge terminated with OH group has the low local work function of 3.76 eV.[16, 17] But the states near the Fermi level have large $E_p$, leading to a large EVBH that is unfavorable to CFE. [18] The case of O terminated Z-edge is in contrary: the states near the Fermi level locate at the gamma point where $E_p$ is vanishing,[19, 20] but $W_0$ is large.[16, 17] The ether group terminated Z-edge possesses both of the advantages: the local work function of 3.96 eV is moderate [9, 16] and the states near the Fermi level locate at the gamma point.[19] The H terminated Z-edge have various of band structures depending on the density of H.[19, 21-25] Nitrogen surface doping is believed to be able to reduce the work function of carbon.[26-29] Partially substituting H with primary amine (NH$_2$) group can alternate the band structure of H terminated Z-edge of graphene, but none of the bottom of the conduction bands locates at the gamma point.[30]

[*] Author to whom correspondence should be addressed. Electronic mail: stslzb@mail.sysu.edu.cn

In the present paper we find a stable Z-edge structure (Fig.1 a, b) where the outermost C atom is replaced with N atom, and then the H atom saturates the N atom. We call this secondary amine (NH) terminated Z-edge. We will show that this structure not just has extraordinary low work function but also has edge states with vanishing $E_p$.

## II. CALCULATION METHOD

The band structure and local work function was calculated via the Vienna Ab inito Simulation Package (VASP).[31] The unit cell is indicated by the dashed box of Fig.1a. The electron-core interactions were treated in the projector augmented wave (PAW) approximation.[32] The density functional is treated by the local density approximation (LDA) (with the Ceperly–Alder exchange correlation potential [33]). A kinetic energy cutoff (400 eV) was used. Full relaxation of magnetization was performed for spin-polarized calculations. All atoms were fully relaxed until the force on each atom is less than 0.01 eV/Å. The vacuum gaps between the graphene ribbons in both parallel and perpendicular directions are 1 nm in the band structure calculation and 3.5 nm in the local work function calculation. The k point mesh is $1\times 50\times 1$ in the band structure calculation and $1\times 10\times 1$ in the local work function calculation.

The electric potential was obtained by integrating the coulomb potential of electrons and ions. We calculated the Perdew-Zunger exchange-and-correlation potential with electron density. These two potentials were summed to give potential barrier. Then it is shifted to let Fermi level equal to zero.

The Mulliken charge calculation is done via Dmol3.[34, 35] The density functional is treated by the LDA (with the PWC exchange correlation potential [36]). Other calculation parameters are the same as the band structure calculation.

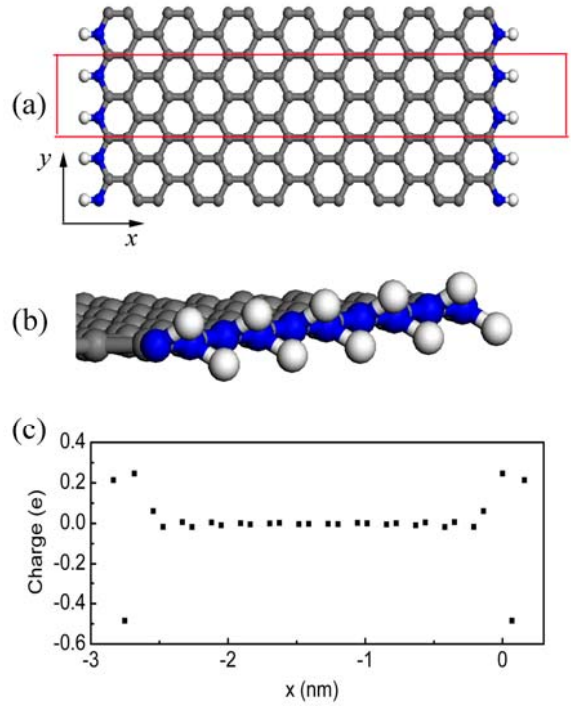

FIG. 1. (Color online) (a)Scheme of the secondary amine (NH) group terminated Z-edge graphene nanoribbon. (b) The stable edge structure after relaxation. (c) Mulliken charge distribution across the graphene nanoribbon.

## III. RESULTS AND DISCUSSION

After relaxation, we found the most stable structure in which the NH groups are connected to the edge carbon with N-H bonds bending up and down from the graphene plane alternatively (Fig.1b). The tilt angle of the C-N bond is $1.0°$. The C-N bond and the N-H bond forms an angle of $115.9°$. The Mulliken charge distribution is presented in Fig.1 (c). One sees that some electrons of the NH groups have been transferred to carbons.

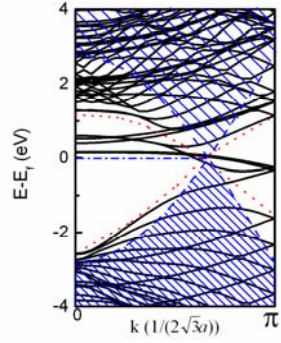

FIG. 2. (Color online) Band structure of the secondary amine group terminated zigzag edge graphene nanoribbon. The extended states of bulk graphene locate in the shadow region. Because of the finite width, the boundaries are shifted to the dotted curves.

The band structure is presented in Fig.2. The bands inside the upper and lower regions bounded by the dotted curves are the extended states of the graphene. There are six bands that penetrate out those regions and extend to the gamma point. In the vicinity of the gamma point these bands become flat asymptotically. They are edge states as shown by Fig.3 where the squares, circles, up-triangles, and down-triangles are the amplitudes of the $s$, $p_y$, $p_z$, and $p_x$ orbitals respectively. The flat bands above neutral Fermi level are roughly extending over two third of the Brillouin zone. Recalling the $y$-directional size of the unit cell is $2\sqrt{3}a$ with $a = 0.142$ $nm$ the bond length, we estimate that the line density of edge states at one edge is $\lambda = 4/(3\times 2\sqrt{3}a) = 2/(3\sqrt{3}a)$. Only the lowest two bands have been considered and the factor 2 of the spin freedoms has been included.

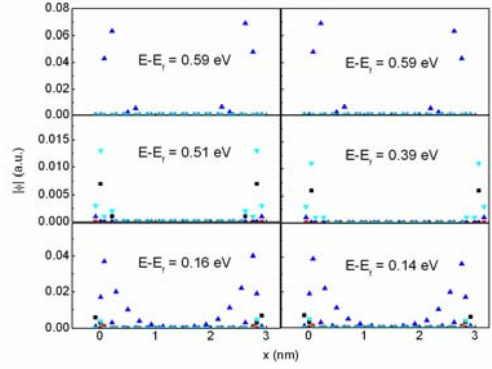

FIG. 3. (Color online) Site projected wave functions of six states near the Fermi level at the gamma point. The squares, circles, up-triangles, and down-triangles are the amplitudes of the *s*, $p_y$, $p_z$, and $p_x$ orbitals respectively.

The highest two flat bands are degenerate at the gamma point. From Fig.3 it is clear that they are edge states formed with the pi-orbitals of carbons, as the tight-binding theory will predict if one neglects the hopping between the s-orbital of nitrogen and the pi-orbital of the carbon in the last column [10]. The other four flat bands are originated from the nitrogen atoms, recalling that our unit cell has four nitrogen atoms. The lowest two flat bands respectively have $E_s = 0.14$ eV and 0.16eV at the gamma point, which are most relevant to the CFE. From the two bottom panels of Fig.3 one sees that they have the *s* orbital as the largest component at nitrogen atoms and the $p_z$ orbital as the dominant component at the carbon atoms in the edge region. Because the graphene ribbon that we calculated has finite width there are artificial negative charges accumulating in the middle of the ribbon that raises the edge potential a little bit. For the half-infinite graphene ribbons, the tight-binding theory predicted that the bands of edge states will be shifted to the Fermi level (the horizontal dashed-dotted line in Fig.2) and will be half occupied in the absent of applied field, while the extended states locate in the shadow region

(Fig.2). [10]

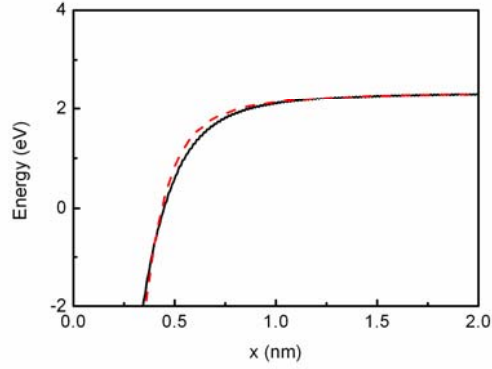

FIG. 4. (Color online) The vacuum potential barrier along the *x*-directional line starting at the last carbon atom (solid) and nitrogen atom (dashed) in the absent of applied fields. The origin of *x* coordinate is at the last carbon atom. The Fermi level is set to zero.

In Fig.4, the solid (dashed) curve is the vacuum potential barrier along the *x*-directional line starting at the last carbon (nitrogen) in the absent of applied fields. The Fermi level has been chosen to be the energy origin. The local work functions for both paths turn out to be the same, equal to $W_0 = 2.30$ eV. The exchange-correlation energy has been included in the potential.[16] When the macroscopic field ($F_0$) is applied to the graphene, electrons will occupy the edge states and screen the applied field. If all edge states were occupied, the induced electrons in the edge region should produce an electric field with magnitude of about $\frac{e\lambda}{2\pi\varepsilon_0 a} = 55$ V/*nm* at the edge, which is much larger than the typical electric fields at the apex of field emitters in the CFE region which are less than 5 V/*nm*. Thus the applied field can be screened via inducing electrons in a fraction of edge states. Because the bands of edge states are flat, the existence of empty edge states implies that the Fermi level is pinned and the EVBH is $W_p = W_0 - E_s + \frac{(\hbar k_y)^2}{2m}$. In other words, the applied fields can be screened almost completely by the induced electrons in the edge states. Therefore, the metallic nano-wall model can be applied to our case. Using the transmission coefficient of the metallic nano-wall [37], we can estimate the emission current line density, resulting

$$J = c_0 e \lambda v e^{-\frac{2bW_p^{5/2}}{5ehF_0^2}} \qquad (1)$$

Where $h$ is the graphene height, $b = 6.83$ eV$^{-3/2}$Vnm$^{-1}$ is the second Fowler-Nordheim constant,[37] $c_0$ is the occupation factor of the edge states which weakly depends on $F_0$ with value between 0.5 and 1, and $v$ is the collision frequency of the electron in the edge state that has the order of $10^{14}$ Hz. [38] From (1) we estimate the turn-on field of 10 V/μm for the current line density of 1 nA/μm from the graphene with $h = 5$ μm. To estimate the emission energy width, we should note that the parallel kinetic energy of the lowest edge band is $E_p = 0.35$ eV at the middle of the Brillouin zone thus the emission current at this point is much smaller than that at the gamma point. The total energy difference between these two points is only 0.03 eV.

## IV. SUMMARY

We found that the zigzag edge of graphene terminated with the secondary amine group has the local work function 2.30 eV that is the lowest local work function of graphene edges that we have known so far. The bands and vacuum potential are calculated via the *ab initio* method. The edge states near the gamma points are found. We show that these edge states pin the Fermi level and are mainly responsible to the cold field emission. The emission current line density is estimated.

## ACKNOWLEDGMENTS

The project was supported by the National Basic Research Program of China (Grant No.

2007CB935500 and 2008AA03A314), the National Natural Science Foundation of China (Grant No. 11104358), the high-performance grid computing platform of Sun Yat-sen University, the Guangdong Province Key Laboratory of Computational Science and the Guangdong Province Computational Science Innovative Research Team.

REFERENCES

[1] G. Eda, H. E. Unalan, N. Rupesinghe, G. A. J. Amaratunga, and M. Chhowalla, Applied Physics Letters 93, 233502 (2008).
[2] S. W. Lee, S. S. Lee, and E. H. Yang, Nanoscale Research Letters 4, 1218 (2009).
[3] A. Malesevic, R. Kemps, A. Vanhulsel, M. P. Chowdhury, A. Volodin, and C. Van Haesendonck, Journal of Applied Physics 104, 084301 (2008).
[4] M. Qian, T. Feng, H. Ding, L. Lin, H. Li, Y. Chen, and Z. Sun, Nanotechnology 20, 425702 (2009).
[5] Z. Shpilman, B. Philosoph, R. Kalish, S. Michaelson, and A. Hoffman, Applied Physics Letters 89, 252114 (2006).
[6] Z. S. Wu, S. F. Pei, W. C. Ren, D. M. Tang, L. B. Gao, B. L. Liu, F. Li, C. Liu, and H. M. Cheng, Advanced Materials 21, 1756 (2009).
[7] W. T. Zheng, Y. M. Ho, H. W. Tian, M. Wen, J. L. Qi, and Y. A. Li, Journal of Physical Chemistry C 113, 9164 (2009).
[8] Z. M. Xiao, J. C. She, S. Z. Deng, Z. K. Tang, Z. B. Li, J. Lu, and N. S. Xu, ACSNano 4, 6332 (2010).
[9] H. Yamaguchi, K. Murakami, G. Eda, T. Fujita, P. Guan, W. Wang, C. Gong, J. Boisse, S. Miller, M. Acik, K. Cho, Y. J. Chabal, M. Chen, F. Wakaya, M. Takai, and M. Chhowalla, Acs Nano 5, 4945 (2011).
[10] D. J. Klein, Chemical Physics Letters 217, 261 (1994).
[11] K. Nakada, M. Fujita, G. Dresselhaus, and M. S. Dresselhaus, Physical Review B 54, 17954 (1996).
[12] M. Fujita, K. Wakabayashi, K. Nakada, and K. Kusakabe, Journal of the Physical Society of Japan 65, 1920 (1996).
[13] K. Wakabayashi, M. Fujita, H. Ajiki, and M. Sigrist, Physical Review B 59, 8271 (1999).
[14] N. M. R. Peres, F. Guinea, and A. H. C. Neto, Physical Review B 73, 125411 (2006).
[15] A. R. Akhmerov and C. W. J. Beenakker, Physical Review B 77, 085423 (2008).
[16] W. L. Wang, J. W. Shao, and Z. B. Li, Chemical Physics Letters 522, 83 (2012).
[17] W. L. Wang and Z. B. Li, Journal of Applied Physics 109, 114308 (2011).
[18] W. L. Wang, X. Z. Qin, N. S. Xu, and Z. B. Li, Journal of Applied Physics 109, 044304 (2011).

[19]   G. Lee and K. Cho, Physical Review B 79, 165440 (2009).
[20]   A. Ramasubramaniam, Physical Review B 81, 245413 (2010).
[21]   Y.-W. Son, M. L. Cohen, and S. G. Louie, Physical Review Letters 97, 216803 (2006).
[22]   T. Wassmann, A. P. Seitsonen, A. M. Saitta, M. Lazzeri, and F. Mauri, Physical Review Letters 101, 096402 (2008).
[23]   J. Kunstmann, C. Ozdogan, A. Quandt, and H. Fehske, Physical Review B 83, 045414 (2011).
[24]   Y. H. Lu, R. Q. Wu, L. Shen, M. Yang, Z. D. Sha, Y. Q. Cai, P. M. He, and Y. P. Feng, Applied Physics Letters 94, 122111 (2009).
[25]   B. Xu, J. Yin, Y. D. Xia, X. G. Wan, K. Jiang, and Z. G. Liu, Applied Physics Letters 96, 163102 (2010).
[26]   J. Xu, J. Mei, X. H. Huang, X. Li, Z. Li, W. Li, and K. Chen, Applied Physics a-Materials Science & Processing 80, 123 (2005).
[27]   W. T. Zheng, C. Q. Sun, and B. K. Tay, Solid State Communications 128, 381 (2003).
[28]   C. F. Shih, K. S. Liu, and I. N. Lin, Diamond and Related Materials 9, 1591 (2000).
[29]   M. Kaukonen, R. M. Nieminen, S. Poykko, and A. P. Seitsonen, Physical Review Letters 83, 5346 (1999).
[30]   F. Cervantes-Sodi, G. Csanyi, S. Piscanec, and A. C. Ferrari, Physical Review B 77, 165427 (2008).
[31]   G. Kresse and J. Furthmuller, Computational Materials Science 6, 15 (1996).
[32]   G. Kresse and D. Joubert, Physical Review B 59, 1758 (1999).
[33]   D. M. Ceperley and B. J. Alder, Physical Review Letters 45, 566 (1980).
[34]   B. Delley, Journal of Chemical Physics 113, 7756 (2000).
[35]   B. Delley, Journal of Chemical Physics 92, 508 (1990).
[36]   J. P. Perdew and W. Yue, Physical Review B 45, 13244 (1992).
[37]   X. Z. Qin, W. L. Wang, N. S. Xu, Z. B. Li, and R. G. Forbes, Proceedings of the Royal Society A 467, 1 (2011).
[38]   J. Peng, Z. B. Li, C. S. He, S. Z. Deng, N. S. Xu, X. Zheng, and G. H. Chen, Physical Review B 72, 235106 (2005).